\documentclass[prb,twocolumn,aps,superscriptaddress,showpacs,floatfix]{revtex4}
\usepackage{amsmath}
\usepackage{graphicx}
\usepackage{color}
\usepackage{tikz}
\begin{document}
\title{Tunneling spectrum of a pinned vortex with a robust Majorana state}

\author{R.S. Akzyanov}
\affiliation{Moscow Institute for Physics and Technology (State
University), 141700 Moscow Region, Russia}
\affiliation{Institute for Theoretical and Applied Electrodynamics, Russian
Academy of Sciences, 125412 Moscow, Russia}

\author{A.V. Rozhkov}
\affiliation{Moscow Institute for Physics and Technology (State
University), 141700 Moscow Region, Russia}
\affiliation{Institute for Theoretical and Applied Electrodynamics, Russian
Academy of Sciences, 125412 Moscow, Russia}
\affiliation{CEMS, RIKEN, Saitama, 351-0198, Japan}

\author{A.L. Rakhmanov}
\affiliation{Moscow Institute for Physics and Technology (State
University), 141700 Moscow Region, Russia}
\affiliation{Institute for Theoretical and Applied Electrodynamics, Russian
Academy of Sciences, 125412 Moscow, Russia}
\affiliation{CEMS, RIKEN, Saitama, 351-0198, Japan}

\author{Franco Nori}
\affiliation{CEMS, RIKEN, Saitama, 351-0198, Japan}
\affiliation{Department of Physics, University of Michigan, Ann Arbor, MI
48109-1040, USA}

\begin{abstract}
We study a heterostructure which consists of a topological insulator and
a superconductor with a hole. The hole pins a vortex. The system supports a
robust Majorana fermion state bound to the vortex core. We investigate the
possibility of using scanning tunneling spectroscopy (i) to detect the
Majorana fermion in the proposed setup and (ii) to study excited states
bound to the vortex core. The Majorana fermion manifests itself as a
magnetic-field dependent zero-bias anomaly of the tunneling conductance.
Optimal parameters for detecting Majorana fermions have been obtained. In
the optimal regime, the Majorana fermion is separated from the excited
states by a substantial gap. The number of zero-energy states equals the
number of flux quanta in the hole; thus, the strength of the zero-bias
anomaly depends on the magnetic field. The lowest energy excitations bound to
the core are also studied.  The excited states spectrum differs from the
spectrum of a typical Abrikosov vortex, providing additional indirect
confirmation of the Majorana state observation.
\end{abstract}


\maketitle

\section{Introduction}

In 1937 Majorana
derived~\cite{Ettore}
an alternative representation of the Dirac equation for particles with spin
1/2. In this representation the Dirac equation has an additional solution,
the so-called Majorana fermion. This unusual particle is equal to its
antiparticle, that is, for the Majorana fermion
\begin{equation}\label{1}
\gamma=\gamma^{\dagger}.
\end{equation}
This is impossible for the usual Dirac fermions. Among the elementary
particles, the neutrino is a candidate for the Majorana fermion, but this
is not firmly established yet. Several
setups~\cite{Wilczek,Wilczek1,
chung_2009,
volovik_2009,
benjamin_2010,
alicea_device,
kraus,
fujimoto_2008,
Sato_Phys.Rev.B_2009,
fu_kane_device,
sau_robustness,
sato,
liang_wang_hu2012,
rozhkov_2kondo}
have been theoretically
proposed for observing Majorana fermions in condensed matter systems, e.g.,
excitations in the quantum Hall effect, in topological superconductors of
$(p_x+ip_y)$-type,
wires with strong spin-orbit interaction, etc. The observation of Majorana
fermions is of interest not only for fundamental physics but also for
potential applications. Majorana fermions are expected to exhibit
non-Abelian statistics and could be used to realize quantum gates that are
topologically protected from local sources of
decoherence~\cite{qcomp}.
Recent
experiments~\cite{frolov,Goldhaber}
hinted at the existence of a Majorana fermion in nanowires coupled to
superconductors and in hybrid superconductor-topological insulator devices.
However, the problem is still open, and no smoking-gun evidence has
surfaced.

\subsection{Previous results}

The interface between a topological insulator and a superconductor is a
candidate system for the possible realization of Majorana
fermions.~\cite{fu_kane_device,sau_robustness,AL,feigel1,feigel2}
Such an interface has been fabricated in
experiments.~\cite{Goldhaber,high-tc_and_ti,morp2011,jjunc_ti,sc_ti_wang2012}
Electrons on the surface of topological insulators are described by the
two-dimensional (2D) massless Dirac equation, where the electron and hole
excitations lie on a Dirac cone
$\varepsilon_{\bf k} = v_{\rm F} |{\bf k}|$,
and the Dirac point of this cone is located at the Fermi level~\cite{top}.
The contact between the topological insulator and the superconductor
generates, through the proximity effect, a finite mass to these Dirac
fermions. In the presence of an external magnetic field, the mass term
acquires a non-trivial complex phase.  Several theoretical proposals for
the realization of Majorana fermions are based on this setup
\cite{fu_kane_device,sau_robustness,AL,feigel1,feigel2}.

\begin{figure}[t!]
\center
\includegraphics [width=8.5cm, height=4.5cm]{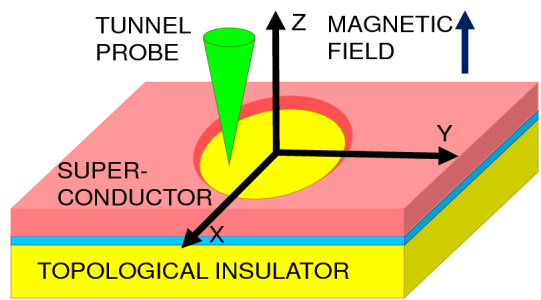}
\caption{(Color online) Proposed experimental setup for detecting a
Majorana fermion. A layer of superconducting material (pink) is
separated from a slab of topological insulator (yellow) by a thin
insulating layer (blue). The external magnetic field is perpendicular to
the interface. A cylindrical hole in the superconductor serves as a pinning
center for a vortex. The tunneling conductance between the tunneling probe
(green) and the open fraction of the topological insulator surface is used
to investigate the low-lying single-electron states bound to the vortex
core.
}
\label{stm_exp}
\end{figure}

In this paper we discuss the system presented in
Ref.~\onlinecite{AL}
(see
Fig.~\ref{stm_exp}).
It consists of three layers: the topological insulator at the bottom, a
sufficiently thick layer of an $s$-wave superconductor on top, and between
them a thin buffer insulating layer, which controls the electron tunneling
between the topological insulator and the superconductor. The system is
placed in a weak magnetic field to create a vortex in the superconductor and,
consequently, a 2D vortex in the 2D superconducting state induced on the
surface of the topological insulator. The core of the 2D vortex hosts the
Majorana fermion
state~\cite{Ivanov_Phys.Rev.Lett._2001}.

In general, a setup of this type has an obvious deficiency, which hampers
the detection of the Majorana fermion state: the minigap separating the
Majorana fermion and the so-called Caroli-de Gennes-Matricon (CdGM) levels
in the core of the Abrikosov
vortex~\cite{zhen}
is too small (about $10^{-2}$~K). To detect the Majorana fermion, both the
temperature and the experimental energy resolution must be smaller than the
minigap, thus, the smallness of the minigap imposes very stringent
requirements on experiments.

Fortunately, it is known that, when an Abrikosov vortex is pinned by a
columnar defect, the minigap increases when increasing the defect
radius~\cite{melnik}.
This happens because the lowest CdGM states are destroyed by the defect.
Consequently, the minigap, as a function of the defect radius $R$,
saturates when
$R \sim \xi$,
where $\xi$ is the superconducting coherence length. In such a regime,
virtually all CdGM states are destroyed. Based on this idea, it was
proposed~\cite{AL}
to pin the Abrikosov vortex on a hollow cylindrical channel in the
superconducting layer (see Fig.1). The purpose of this ``hole" is two-fold:
It rids the system of a large set of parasitic CdGM excitations, and also
allows access to the surface of the topological insulator.

As for CdGM states inside the 2D vortex core, it was demonstrated
\cite{sau_robustness,AL}
that, if the chemical potential of the 2D Dirac electrons lies sufficiently
close to the Dirac point, the corresponding minigap is quite large, making
the Majorana state ``robust" (related ideas for different physical systems
have been discussed in
Ref.~\onlinecite{sato}).

\subsection{Our results}

The above arguments, however, are purely theoretical. To demonstrate that
the proposed system does indeed hosts a Majorana fermion, a reliable
experimental proof is required. The purpose of this paper is to investigate
the usefulness of scanning tunneling spectroscopy (STS) as a tool to
diagnose the presence of the Majorana fermion in the setup of Fig.~1.

To address this question we discuss two related problems: First, what STS
features are associated with the presence of the Majorana fermion in our
setup; second, what are the system parameters which optimize the
observation of these features. Below we will demonstrate that, even at
not-too-low temperatures, the tunneling spectrum can be used to identify
the Majorana state in the proposed system.

The remainder of the paper is organized as follows.
In
Sec.~\ref{bdg}
the Bogolyubov-de Gennes equations are derived.
The zero-energy solutions (zero modes) of these equations are discussed in
Sec.~\ref{zero_modes}.
The tunneling spectrum at arbitrary energy is discussed in
Sec.~\ref{tunn_spectrum}.
The results are discussed in
Sec.~\ref{discussion}.

\section{Bogolyubov-de Gennes equations}
\label{bdg}

In this section we derive the differential equations for the wave functions
of the single-electron eigenstates bound to the vortex core. The proximity
effect in 2D materials has been studied in several papers
\cite{volkov1995,kopnin_melnikov,kopnin_khaymovich}.
Our derivation generalizes the procedure of
Refs.~\onlinecite{sau_robustness,AL}
to account for an arbitrary number of vortices trapped in the hole. The
presentation below is quite sketchy. For extra details the reader should
consult
Refs.~\onlinecite{sau_robustness,AL}.

\subsection{Microscopic model}

The Hamiltonian of the system can be written
as~\cite{sau_robustness}
\begin{equation}\label{H}
 H=H_{\rm TI}+H_{\rm SC}+{T} +{T}^\dagger,
 \end{equation}
where
$H_{\rm TI},H_{\rm SC}$
are related to the topological insulator (TI) and the superconductor (SC),
respectively. The term $T$ describes the tunneling from the topological
insulator to the superconductor, and
$T^\dagger$
accounts for the tunneling from the superconductor to the topological
insulator. The corresponding Bogolyubov-de Gennes equations
\cite{zhen}
are
($\hbar = 1$)
\begin{equation}\label{H_ti}
H_{\rm TI}{\psi}_{\rm TI}+T^\dagger {\psi}_{\rm SC}=\omega{\psi}_{\rm TI},
\end{equation}
\begin{equation}\label{H_sc}
 H_{\rm SC}{\psi}_{\rm SC}+T {\psi}_{\rm TI}=\omega{\psi}_{\rm SC}.
 \end{equation}
The terms
$H_{\rm TI},H_{\rm SC}$
can be written as
$4\times4$
matrices  in the Nambu basis,
\begin{eqnarray}
\label{Nambu}
\nonumber
  H_{\rm TI} &=& [iv(\sigma\cdot\nabla_r)-U(\textbf{r})]\tau_z, \\
  H_{\rm SC} &=&
		-\left(
			E_{\rm F} + \frac {{\nabla}_{\textbf{R}}^2}{2m}
		 \right)
		\tau_z
		+
		\Delta'(\textbf{R})\tau_x+\Delta''(\textbf{R})\tau_y,
\end{eqnarray}
and
$T=\tau_z{\cal T}(\mathbf{R-r})$.
In these equations,
$\textbf{R}=(x,y,z)$
is a point in the bulk of the superconductor,
$\textbf{r}=(x,y)$
is a point on the surface of the topological insulator,
$\sigma_j,\tau_j$
are the spin and charge Pauli matrices,
$\Delta', \Delta''$
are the real and imaginary parts of the order parameter in the
superconductor, $v$ is the Fermi velocity of the electrons on the surface
of the topological insulator,
$E_{\rm F}$
is the Fermi energy in the superconductor, and
$U(r)$
is a gate voltage applied to control the Fermi level in the
topological insulator.~\cite{sau_robustness}
The wave functions
$\psi_{\rm TI,\,SC}$
are four-component spinors:
\begin{eqnarray}
\psi_{\rm TI,\,SC}=[u_\uparrow,u_\downarrow,v_\downarrow,-v_\uparrow]^T.
\end{eqnarray}

In Hamiltonian
Eq.~(\ref{Nambu})
the vector-potential describing the magnetic field is omitted. This is
justified provided that the flux, passing through the area where the subgap
wave functions are localized, is smaller than the flux quantum. In the
regime we study, the subgap states is localized within distance
$r \sim \xi$
from the hole center, consequently, the magnetic field may be neglected
when
$(\xi/\lambda_{\rm L})^2 \ll 1$,
where
$\lambda_{\rm L}$
is the London penetration depth in the superconducting film. Thus, for the 
type-II superconducting film this condition is well satisfied.

We also neglect the effects of the magnetic field on the superconductor.
The magnetic field necessary for a flux quantum to enter the
superconductor, $H_{c1}$, is much smaller than the thermodynamic field
$H_c$. Thus, the effects of magnetic field on the superconductor are
expected to be quite moderate.

It is easy to check that $H$ satisfies the following charge-conjugation
symmetry:
\begin{eqnarray}
H = - \tau_y \sigma_y H^* \tau_y \sigma_y.
\label{charge_conj}
\end{eqnarray}
Consequently, for every eigenstate
$\psi$
of $H$ with a nonzero eigenenergy
$\omega \ne 0$,
an eigenstate
$\tau_y \sigma_y \psi^*$
with eigenenergy $-\omega$ is present. This symmetry is very robust:
Disorder potential does not destroy this property.

\subsection{Effective Hamiltonian}

Following
Ref.~\onlinecite{sau_robustness}
we exclude
$\psi_{\rm SC}$
from
Eqs.~\eqref{H_ti}
and
\eqref{H_sc}
to derive
\begin{eqnarray}
\label{bdg_eq}
(H_{\rm TI} + \Sigma )\psi_{\rm TI}=\omega\psi_{\rm TI},
\\
\Sigma = T^\dagger(\omega - H_{\rm SC})^{-1}T.
\end{eqnarray}
We are interested in bound states with energies lying within the
superconducting energy gap
$|\omega|<|\Delta|$.
In this case, the self-energy matrix $\Sigma$ can be calculated quite
straightforwardly.
\cite{sau_robustness,AL}
For low-lying electron states
${\bf k} \approx {\bf M}$
(here
${\bf M}$
is the location of the Dirac cone apex in the topological insulator
Brillouin zone), it is equal to
\begin{eqnarray}
\label{Sigma}
\Sigma_{{\bf M}, \omega}
=
\lambda
\frac{
	\Delta \tau_x - \omega \tau_0
     }
     {
	\sqrt{|\Delta|^2 - \omega^2}
     }
-
\delta U \tau_z,
\end{eqnarray}
where
$\tau_0$
is the
$2\times2$
identity matrix. The parameter $\lambda$ has the dimension of energy. It
characterizes the transparency of the barrier between the topological
insulator and the superconductor:
\cite{sau_robustness}:
When
$\lambda \sim  E_{\rm F}$
($\lambda \ll  E_{\rm F}$),
the barrier is transparent (non-transparent). The parameter
$\delta U=O(\lambda)$
is the shift of the topological insulator chemical potential due to doping by the
superconductor.

Using
Eq~(\ref{Sigma})
we can cast the Bogolyubov-de Gennes equation
(\ref{bdg_eq})
in the form
\begin{eqnarray}
\label{bdg2}
H_{\rm eff} \psi_{\rm TI} = \omega \psi_{\rm TI},
\end{eqnarray}
where the effective Hamiltonian
$H_{\rm eff}$
and its parameters
are~\cite{sau_robustness}
\begin{equation}\label{Heff}
H_{\rm eff}=[i\tilde{v}(\omega)(\sigma \cdot \nabla_{\textbf{r}})-\tilde{U}(\omega)]\tau_z+\tilde{\Delta}'(\omega)\tau_x+\tilde{\Delta}''(\omega)\tau_y,
\end{equation}
\begin{equation}
\tilde{v}(\omega)=\frac {v\sqrt{|\Delta|^2-\omega^2}}{\sqrt{|\Delta|^2-\omega^2}+\lambda},
\end{equation}
\begin{equation}
\tilde{U}(\omega)=\frac {(U + \delta U)
\sqrt{|\Delta|^2-\omega^2}}{\sqrt{|\Delta|^2-\omega^2}+\lambda},
\end{equation}
\begin{equation}\label{teta}
\tilde{\Delta}(\omega)=\frac {\Delta \lambda}{\sqrt{|\Delta|^2-\omega^2}+\lambda}.
\end{equation}
We see that the effective parameters experience energy-dependent
renormalization with respect to the bare quantities.

\subsection{Normalization of the effective wave function}

In addition to the effective Hamiltonian, it is desirable to have a
normalization condition for the effective wave function
$\psi_{\rm TI}$.
The normalization condition in the
$\textbf{k}$-space
for the full wave function is
\begin{eqnarray}
\label{norm0}
\int_{\bf k}
	(\psi_{\rm TI}^{{\bf k},\omega})^\dag
	\psi_{\rm TI}^{{\bf k}, \omega}
+
\int_{{\bf k} k_z}
	(\psi_{\rm SC}^{{\bf k},k_z, \omega})^\dag
	\psi_{\rm SC}^{{\bf k},k_z, \omega }
=
1,
\end{eqnarray}
where the symbol
$\int_{\bf k}$
stands for
$\int {d^2 {\bf k}}/{(2\pi)^2}$,
and
$\int_{{\bf k}, k_z}$
stands for
$\int {d^2 {\bf k} dk_z}/{(2\pi)^3}$.
 Excluding
$\psi_{\rm SC}^{{\bf k},k_z, \omega}$,
we can rewrite the latter equation as
\begin{eqnarray}
\int_{\bf k}
	(\psi_{\rm TI}^{{\bf k}, \omega})^\dag
	\psi_{\rm TI}^{{\bf k}, \omega}
+
\int_{\bf k}
	(\psi_{\rm TI}^{{\bf k}, \omega})^\dag
	{\widehat P}_{{\bf k}, \omega}
	\psi_{\rm TI}^{{\bf k}, \omega}
= 1,
{\rm \ \ where}
\\
{\widehat P}_{{\bf k}, \omega}
=
\int_{k_z}
T^\dag_{{\bf k}, k_z}
	( \omega - H_{\rm SC}^{{\bf k}, k_z} )^{-2}
T_{{\bf k}, k_z}
=
- \frac{\partial \Sigma_{{\bf k}, \omega}}{\partial \omega}.
\end{eqnarray}
We will see below that
$\psi_{\rm TI} ({\bf r})$
varies over a length scale
$\sim \xi$.
Consequently,
$v|{\bf k - M}| \sim \Delta$.
In such a regime, we can assume that
${\widehat P}_{{\bf k}, \omega}
\approx
{\widehat P}_{{\bf M}, \omega}$.
Using
Eq.~(\ref{Sigma}),
where
$\delta U$
is virtually independent of $\omega$, we obtain
\begin{eqnarray}
{\widehat P}_{{\bf M}, \omega}
=
\lambda \Delta
\frac{\Delta\tau_0 - \omega \tau_x}{(\Delta^2 - \omega^2)^{3/2}},
\qquad |\omega| < \Delta.
\end{eqnarray}
In this approximation
${\widehat P}$
is momentum-independent, and
Eq.~(\ref{norm0})
can be rewritten in real space as
\begin{eqnarray}
\label{norm_ti}
\int d^2{\bf r}
	[\psi_{\rm TI}^\omega ({\bf r})]^\dag
	(1 + {\widehat P}_{{\bf M}, \omega})
	\psi_{\rm TI}^\omega ({\bf r})
= 1.
\end{eqnarray}
Observe that for
$|\omega|$
approaching
$|\Delta|$,
the matrix
$\widehat P$
diverges. This divergence occurs because in the regime
$0< |\Delta| - |\omega| \ll |\Delta|$
an electron spends a large portion of its time in the superconductor.
Therefore, the norm of
$\psi_{\rm SC}=\widehat P \psi_{\rm TI}$
increases relative to the norm of
$\psi_{\rm TI}$.

\subsection{Equations for the effective wave function}

We are looking for solutions of the Bogolyubov-de Gennes equations
Eq.~(\ref{bdg2})
which correspond to bound states. Consequently, the energies of these
solutions $\omega$ should be smaller than the proximity-induced gap
$\Delta_{\rm TI}$,
which satisfies the
equation~\cite{AL}
\begin{equation}\label{DTI}
\frac{\Delta_{\rm TI}}{\lambda}
=
\sqrt{\frac{\Delta-\Delta_{\rm TI}}{\Delta+\Delta_{\rm TI}}}.
\end{equation}
Imagine now that $l$ vortices end up trapped by the hole. In such a
situation, the order parameter
$\Delta(\textbf{r})$
can be expressed
as~\cite{tinkham}
\begin{equation}
 \Delta(\textbf{r})=|\Delta(r)|\exp(-il\theta),
\end{equation}
where $r$ and $\theta$ are polar coordinates, and
$|\Delta(r)|\rightarrow |\Delta|$
when
$r\rightarrow \infty$.
If the hole radius $R$ is large, $R>\xi$, $|\Delta(r)|$ can be approximated
as
\begin{equation}
\label{abs_Delta}
|\Delta(r)|=|\Delta|\Theta(r-R),
\end{equation}
where $\Theta(r)$ is the Heaviside step function.

Let us define a spinor $F$ as
\begin{eqnarray}\label{F}
\nonumber
\psi_{\rm TI} &=& \exp[-i\theta(l\tau_z+\sigma_z)/2+i\mu \theta]F^{\mu}(r), \\
F^{\mu} &=& (f^{\mu}_1,f^{\mu}_2,f^{\mu}_3,-f^{\mu}_4)^T.
\end{eqnarray}
The physical meaning of $\mu$ is the total angular momentum of the state.
The transformation in
Eq.~(\ref{F})
is well-defined only when
\begin{eqnarray}
\label{mu_cond}
j = \mu + \frac{l + 1}{2}
\end{eqnarray}
is an integer. In other words, when the number of vortices $l$ is odd
(even), the angular momentum $\mu$ is integer (half-integer).

Substituting Eqs.~\eqref{Heff}, \eqref{teta}, and \eqref{F} in
Eq.~(\ref{bdg2})
we derive
\begin{eqnarray}\label{final}
\nonumber
i\tilde{v}\!\left (\frac d{dr}+\frac {2\mu+l+1}{2r} \right
)\!f^{\mu}_2+|\tilde{\Delta}|f^{\mu}_3-(\omega\!+\!\tilde{U})f^{\mu}_1=0,
\\ \nonumber
i\tilde{v}\!\left (\frac d{dr}-\frac {2\mu+l-1}{2r} \right
)\!f^{\mu}_1-|\tilde{\Delta}|f^{\mu}_4-(\omega\!+\!\tilde{U})f^{\mu}_2=0,
\\
i\tilde{v}\!\left (\frac d{dr}+\frac {2\mu-l+1}{2r} \right
)\!f^{\mu}_4+|\tilde{\Delta}|f^{\mu}_1-(\omega\!-\!\tilde{U})f^{\mu}_3=0,
\\
\nonumber
i\tilde{v}\!\left (\frac d{dr}-\frac {2\mu-l-1}{2r} \right
)\!f^{\mu}_3-|\tilde{\Delta}|f^{\mu}_2-(\omega\!-\!\tilde{U})f^{\mu}_4=0.
\end{eqnarray}
These equations are the foundation on which the main results of this paper
are based. These equations will be solved and analyzed for different values
of $\omega$, $\mu$, and $l$. Since
Eqs.~(\ref{final})
admits the following symmetry:
$\mu \leftrightarrow - \mu$,
$f_4\leftrightarrow if_1$,
$f_3 \leftrightarrow if_2$,
$\tilde{U} \leftrightarrow -\tilde{U}$,
only $\mu\geq0$ solutions have to be found explicitly.

We mentioned above that, upon contact, the superconductor dopes the surface
states of the topological insulator. Consequently, $U$ becomes a function
of $r$. However, we
assume below that
$U = 0$,
since this condition is most favorable for the observation of the Majorana
fermion. To satisfy this requirement, an external gate electrode
controlling $U$ might be necessary. If
$U(r)$
is nonzero, yet remains small for any $r$, then perturbation theory can be
used to account for it.

\section{Zero-energy solution}
\label{zero_modes}

In this section, we will obtain all zero-energy ($\omega=0$) solutions.
Such solutions are often called ``zero modes".  It will be shown that the
number of zero modes is equal to the number of vortices in the hole $l$.

If
$\omega=\tilde{U}=0$,
the system of Eq.
\eqref{final}
decouples into two sets of equations
\begin{eqnarray}\label{1_4}
\nonumber
i\tilde{v}\!
\left(
	\frac d{dr}-\frac {2\mu+l-1}{2r}
\right)\!f^{\mu}_1-|\tilde{\Delta}|f^{\mu}_4=0,
\\
i\tilde{v}\!
\left(
	\frac d{dr}+\frac {2\mu-l+1}{2r}
\right)\!f^{\mu}_4+|\tilde{\Delta}|f^{\mu}_1=0,
\end{eqnarray}
and
\begin{eqnarray}\label{2_3}
\nonumber
i\tilde{v}\!\left (\frac d{dr}+\frac {2\mu+l+1}{2r} \right )\!f^{\mu}_2+|\tilde{\Delta}|f^{\mu}_3=0, \\
i\tilde{v}\!\left (\frac d{dr}-\frac {2\mu-l-1}{2r} \right )\!f^{\mu}_3-|\tilde{\Delta}|f^{\mu}_2=0,
\end{eqnarray}
where
$| \tilde \Delta| = | \tilde \Delta (r)|$
is given by
Eq.~(\ref{abs_Delta}).

Outside the hole ($r>R$), the gap
$| \tilde \Delta |$
is nonzero. Finite solutions of
Eqs.~\eqref{1_4}
and
\eqref{2_3}
can be expressed in terms of the modified Bessel functions $K_m(x)$:
\begin{eqnarray}
\nonumber
f_1\! =\! Ar^{\frac{l}{2}}K_{\mu-1/2}\!\!
\left(\frac{\lambda r}{v}\right),
\,\,
f_4 \!=\! iAr^{\frac{l}{2}}K_{\mu+1/2}\!\!
\left(\frac{\lambda r}{v}\right),
\\
\label{r>R}
f_2\! =\! Br^{-\frac{l}{2}}K_{\mu+1/2}\!\!
\left(\frac{\lambda r}{v}\right),
\,\,
 f_3 \!=\! iBr^{-\frac{l}{2}}K_{\mu-1/2}\!\!
\left(\frac{\lambda r}{v}\right).
\end{eqnarray}
In the hole ($r<R$) we have
$\tilde{\Delta}=0$
and these systems decouple further into four independent equations. They
can be easily solved:
\begin{eqnarray}\label{r<R}
\nonumber
f_1=C_1r^{\mu+\frac{l-1}2},
\quad\ \ \
f_4=C_4r^{\frac {l-1}2 - \mu},
\\
f_2=C_2r^{-\mu-\frac{l+1}2},
\quad\ \ \
f_3=C_3r^{\mu-\frac {l+1}2}.
\end{eqnarray}
Of these four functions,
$f_2$
has the strongest singularity at
$r=0$.
Since a wave function has to be normalizable:
$
\int rdr |f_2|^2 < \infty,
$
the divergence of
$f_2$
must not be too strong:
$\mu + (l+1)/2 < 1$.
For positive $\mu$ and $l$ this inequality cannot be satisfied
simultaneously with the
condition~(\ref{mu_cond}).
Therefore,
$C_2 = 0$.
Further, the function
$f_4$
is normalizable when
\begin{eqnarray}
\label{zero_mode_cond}
\mu < \frac{l+1}{2}.
\end{eqnarray}
Matching the solutions of Eqs.~(\ref{r<R}),
and~(\ref{r>R})
at
$r=R$,
we conclude that
$f_2=f_3=0$,
while
$f_{1,4}$ are non-zero only if
Eq.~(\ref{zero_mode_cond})
is satisfied.

Using the symmetry between positive and negative $\mu$, one can generalize
Eq.~(\ref{zero_mode_cond})
for arbitrary $\mu$:
\begin{equation}
\label{condit}
|\mu| < \frac{l+1}{2}.
\end{equation}
As explained above, the condition that $j$,
Eq.~(\ref{mu_cond}),
is an integer implies that $\mu$ is an integer, if $l$ is odd, and $\mu$ is
a half-integer, if $l$ is even. Keeping this in mind, one discovers that
there are no zero-energy solutions in the absence of the vortex, $l=0$.
There is a single zero mode with $\mu=0$, if $l=1$. In the case of two
vortices in the hole, $l=2$, we have two zero-energy solutions with
$\mu=\pm 1/2$; if $l=3$ there exist three zero-energy solutions  with
$\mu=0,\pm1$,
etc. One can convince oneself that the number of zero-energy solutions
coincides with the number $l$ of vortices in the hole.

This connection between the number of zero modes and $l$ may be detected
experimentally: It implies that the zero-bias anomaly of the tunneling
spectrum is sensitive to the magnetic field. We will discuss this in more
detail in
Sec.~\ref{discussion}.


\section{System with a single vortex}
\label{tunn_spectrum}

In this section we study a system with a single vortex pinned by a hole
($l = 1$).
Since $l$ is odd, the vortex hosts a single Majorana fermion. This Majorana
fermion state can be detected in the tunneling experiment depicted in
Fig.~\ref{stm_exp}.
It manifests itself as a zero-bias anomaly of the tunneling spectrum. We
will determine the parameter range where the zero-bias anomaly current is
the strongest.

In addition to the Majorana fermion state, a set of subgap excited states
is localized in the core of the vortex (the term ``subgap state" implies
that the
eigenenergy of such a state lies within the bulk single-electron gap:
$|\omega| < \Delta_{\rm TI}$).
Unlike a typical Abrikosov vortex, whose core is filled with a dense CdGM
spectrum, in our situation the number of subgap states is small: There
can be as few as two states with positive eigenenergies (and, respectively,
two states with $\omega<0$). We will numerically calculate the subgap
spectrum and discuss the optimization of the system parameters to
facilitate the detection of this spectrum in a tunneling experiment.

\subsection{Majorana fermion}

When
$l=1$
the solution for the Majorana fermion,
Eqs.~\eqref{r>R} and \eqref{r<R} with
$\mu = 0$,
continuous at
$r=R$,
reads
\begin{eqnarray}
\label{Majorana fermion1}
\nonumber
f_1=-i f_4=C_1,\qquad r<R,  \\
f_1=-i f_4=C_1\exp{\left[-\lambda(r-R)/v\right]},\qquad r>R,
\end{eqnarray}
where $C_1$ is a constant.

For the tunneling experiment depicted in Fig.~1, it is important that the
wave function of a probed state is well-localized within the hole. As a
measure of such localization, let us calculate the following ratio:
\begin{eqnarray}
I
=
\frac{
	\int\limits_R^{+\infty}\!
		(1+{\widehat P}_{{\bf M}, 0}) \rho(r)2\pi r\, dr
     }
     {
	\int\limits_0^R \!\rho(r)2\pi r\, dr
     },
\end{eqnarray}
where the probability density $\rho(r)$ is equal to
\begin{eqnarray}
\label{Nb}
\rho(r)
&=&|f_1(r)|^2 +
|f_2(r)|^2 +
|f_3(r)|^2 +
|f_4(r)|^2
\quad
\\
\nonumber
&=&2|C_1|^2\left\{
  \begin{array}{ll}
    1, & r<R, \\
    \exp{\left[-\frac{2\lambda}{\Delta}\frac{(r-R)}{\xi}\right]}, & r>R.
  \end{array}
\right.
\end{eqnarray}
and the operator
${\widehat P}_{{\bf M}, 0}$
(accounts for the tunneling into the superconductor) equals to:
\begin{equation}
{\widehat P}_{{\bf M}, 0}
=
\left\{
  \begin{array}{ll}
    0, & r<R, \\
    \frac{\lambda}{\Delta}, & r>R.
  \end{array}
\right.
\end{equation}
Here we use the relation
$v=\Delta\xi$.
The quantity $I$ varies from $0$ to
$+\infty$.
If
$I=0$,
the Majorana fermion is localized entirely within the hole radius; when $I$ is large the
wave function spreads out deeply into the bulk. Thus, to enlarge the
tunneling current we want to have a small $I$. Simple calculations show
that
\begin{equation}\label{ratio}
I
=
\frac{1}{2}\!
\left(
	\frac{\xi}{R}
\right)^2\!\!
\left(
	\frac{\Delta}{\lambda}
\right)^2\!\!
\left(
	1\!+\!\frac{2\lambda R}{\Delta\xi}
\right)
\left(
	1\!+\!\frac{\lambda}{\Delta}
\right).
\end{equation}
Since $R>\xi$, to have a small $I<2$, we need
$\lambda/\Delta>1.7$.
Below we will see that this inequality will be satisfied in the optimal
regime.

\subsection{Equation for the energies of the excited states}

When
$\omega \ne 0$,
the solution of
Eqs.~\eqref{final}
in the hole
($r<R$)
can be expressed in terms of the Bessel functions
$J_\nu (z)$:
 \begin{eqnarray}
\nonumber
f_1^{\mu}=iAJ_{\mu}\!\!\left(\frac {\omega}{\Delta}\frac {r}{\xi}\right),
\quad
f_2^{\mu}=AJ_{\mu+1}\!\!\left(\frac {\omega}{\Delta}\frac r{\xi}\right),
\\
f_3^{\mu}=iBJ_{\mu-1}\!\!\left(\frac {\omega}{\Delta}\frac r{\xi}\right),
\quad
f_4^{\mu}=BJ_{\mu}\!\!\left(\frac {\omega}{\Delta}\frac r{\xi}\right).
\end{eqnarray}
If $r>R$, it is convenient to introduce the following linear combinations
\cite{AL}
\begin{eqnarray}\label{Center}
 \nonumber
 X_1^{\mu}=if^{\mu}_1+f^{\mu}_4,\quad X_2^{\mu}=if^{\mu}_1-f^{\mu}_4,  \\
 Y_1^{\mu}=if^{\mu}_2+f^{\mu}_3,\quad Y_2^{\mu}=if^{\mu}_2-f^{\mu}_3,
\end{eqnarray}
\begin{eqnarray}
\nonumber
Y_1^{\mu}=\frac {i\tilde{v}}{\omega}\left(\frac{dX_1^{\mu}}{dr}-\frac{1}{\tilde{\xi}}X_1^{\mu}-\frac 1{r}X_2^{\mu}\right), \\
Y_2^{\mu}=\frac {i\tilde{v}}{\omega} \left(\frac {dX_2^{\mu}}{dr} +\frac {1}{\tilde{\xi}}X_2^{\mu}-\frac 1{r}X_1^{\mu}\right),
\\
{\rm where\ \ }
\tilde{\xi}({\omega})
=
\frac {\tilde{v}}{\sqrt {|\tilde{\Delta}|^2-{\omega}^2}},
\end{eqnarray}
and express the solutions in terms of Whittaker functions~\cite{abr}
\begin{equation}
X^{\mu}_{1,2}
=
\frac {C_{1,2}}{\sqrt{r}}\;
W_{\alpha_{1,2},\mu} \! \left(\frac {2r}{\tilde{\xi}(\omega)}\right ),
\end{equation}
\begin{equation}
\alpha_{1,2}
=
\mp \frac {|\tilde{\Delta}|}{2\sqrt {|{\tilde{\Delta}}|^2-{\omega}^2}}.
\end{equation}
Since we seek the subgap solutions
($\omega<|\Delta_{\rm TI}|$),
the values
$\alpha_{1,2}(\omega)$
and
$\tilde{\xi}(\omega)$
are real, and the latter can be considered as a characteristic localization
length of the excitation with energy $\omega$. Matching solutions at
$r=R$,
we derive the following equation for the eigenenergies $\omega$ of the
subgap excited states
\begin{eqnarray}\label{Energy_eq}
\left(\frac {W'_{\alpha_1,{\mu}}}{\tilde{\xi} W_{\alpha_1,{\mu}}}+\frac
{W'_{\alpha_2,{\mu}}}{\tilde{\xi} W_{\alpha_2,{\mu}}}-\frac
{{\mu}+1/2}R+\frac {\omega J_{\mu+1}}{\tilde{v}_{\mu} J_{\mu}}\right)
\nonumber
\\
\times   \left (\frac {W'_{\alpha_1,{\mu}}}{\tilde{\xi}
W_{\alpha_1,{\mu}}}+\frac {W'_{\alpha_2,{\mu}}}{\tilde{\xi}
W_{\alpha_2,{\mu}}}+\frac {{\mu}-1/2}R-\frac { \omega J_{\mu-1}}{\tilde{v}_{\mu} J_{\mu}}\right)=\nonumber\\=\left( \frac {W'_{\alpha_1,{\mu}}}{\tilde{\xi} W_{\alpha_1,{\mu}}}-\frac {W'_{\alpha_2,{\mu}}}{\tilde{\xi} W_{\alpha_2,{\mu}}}-\frac {\tilde{\Delta}}{\tilde{v}}\right)^2 .
\end{eqnarray}
Here the Whittaker functions
$W_{\alpha,\mu}(z)$
are taken at
$z=2R/\tilde{\xi}(\omega)$
and the Bessel functions
$J_\alpha(z)$
at
$z=\omega R/v$.
Prime means differentiation over $z$:
$W'_{\alpha,\mu} (z) = d W_{\alpha,\mu} (z) / dz$.

Equation~(\ref{Energy_eq})
corrects a misprint in Eq.~(41) of
Ref.~\onlinecite{AL}.
There, instead of the valid
$(\mu \pm 1/2)/R$
terms, the incorrect
$(\mu \pm 1)/R$
are shown.

\subsection{The first and higher excited states}

\begin{figure}[t!]
\center
\includegraphics [width=8.5 cm, height=7 cm]{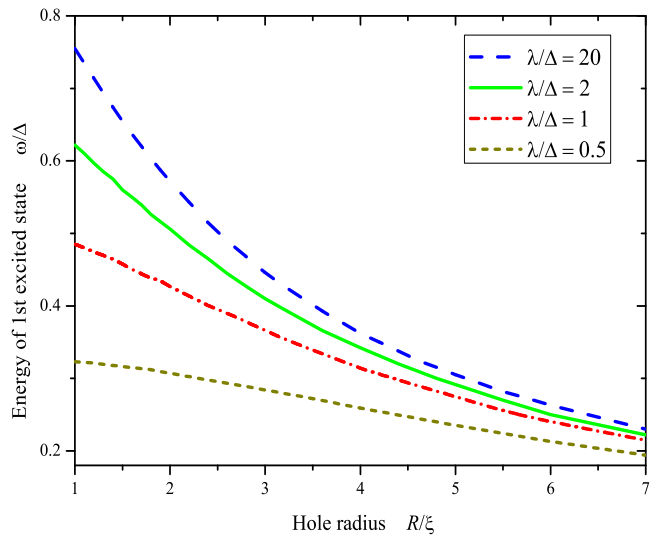}
\caption{(Color online) Normalized energy of the first excited state
($\mu=1$
and
$n=0$)
as a function of the normalized hole radius
$R/\xi$
for different barrier transparencies
$\lambda$.
The energy of the first excited state is bounded from above by
$\Delta_{\rm TI}(\lambda)$
(the gap in the topological insulator), which is a decreasing function of
$\lambda$.
For
$\lambda$=20$\Delta$,
solving Eq.~(\ref{DTI}),
we find that
$\Delta_{\rm TI} \approx \Delta$.
When
$\lambda$=2$\Delta$,
the gap
$\Delta_{\rm TI} \approx 0.75\Delta$;
when
$\lambda$=$\Delta$,
the gap
$\Delta_{\rm TI} \approx 0.54\Delta$.
Finally,
$\Delta_{\rm TI}(0.5 \Delta ) \approx 0.35\Delta$.
}
\label{first_ex}
\end{figure}

Equation~(\ref{Energy_eq})
can be used to study the dependence of the eigenenergies of the subgap
states on the system parameters
$R/\xi$
and
$\lambda/\Delta$.
Each excited state can be
characterized~\cite{AL,zhen}
by a pair of quantum numbers
$\mu,n$,
where $n$ is the principal quantum number of a solution of
Eq.~\eqref{Energy_eq}
with a given $\mu$.

Our numerical analysis shows that the lowest excited state of our system
corresponds to the quantum numbers
$\mu=1$ and $n=0$.
The energy of the first excited state, as a function of the hole radius
$R > \xi$,
is plotted in
Fig.~\ref{first_ex}
for different barrier transparencies
$\lambda/\Delta$.
Note that we do not calculate the energy for small values of $R$. Indeed,
if
$R<\xi$,
the developed formalism becomes invalid and, in addition, in such a regime
the Caroli-de Gennes-Matricon levels begin populating the core of the
vortex.

As we can see from
Fig.~\ref{first_ex},
the energy gap between the first excitation and the Majorana fermion
decreases when $R$ increases. This is quite a natural behavior: The growth
of the radius $R$ leads to an increase of the effective confinement area.
As it is seen from the results shown in
Fig.~\ref{first_ex},
the hole radius must not exceed several $\xi$, otherwise, the gap between
the Majorana fermion and the exited states shrinks too much. It also
follows from
Fig.~\ref{first_ex}
that the increased transparency of the barrier between the topological
insulator and the superconductor,
$\lambda\gg\Delta$,
does not give rise to a significant increase of the gap compared with the
case
$\lambda/\Delta \simeq 2$.
If we choose
$2 < R/\xi < 4$
and
$\lambda \geq 2\Delta$,
then the gap is about
0.4--0.6
in units of
$\Delta$.

Similar to
Eq.~(\ref{Nb}),
we can calculate the probability density
$\rho_1(0)$
in the center of the hole for the first excited state. The corresponding
wave function is given by
Eqs.~\eqref{r>R}
and~\eqref{r<R}.
These expressions have to be matched at
$r=R$.
For
$r>R$,
the normalization condition
Eq.~(\ref{norm_ti})
must be used. Numerical results show that
$\rho_1(0)$
for the chosen range of parameters is of the same order as that for the
Majorana fermion
$\rho(0)$:
$\rho_1(0)/\rho(0)\approx 0.67$.
The excitation with the orbital number $-\mu$ has the same energy with the excitation $\mu$.
Total density of the states at the center of the hole with the same energy $\omega_1$ would be $2\rho_1(0)$:
$2\rho_1(0)/\rho(0)\approx 1.3$.
This means that, in an idealized tunneling experiment, both states manifest
themselves as peaks of comparable magnitude.

\begin{figure}[t!]
\center
\includegraphics [width=8.5 cm, height=7 cm]{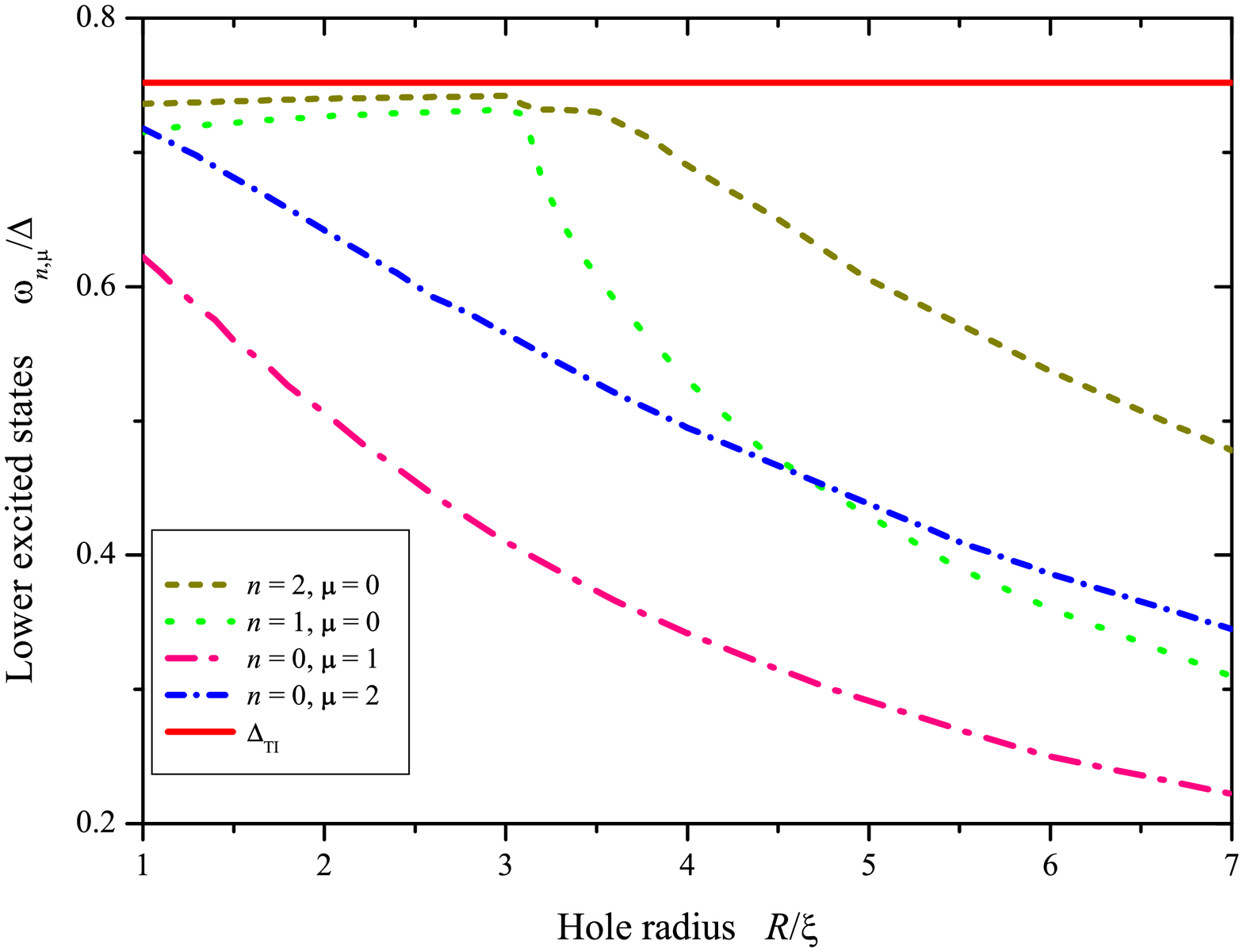}
\caption{(Color online) Energy of the low-lying excited states as a
function of the normalized hole radius
$R/\xi$
for the barrier transparency
$\lambda = 2 \Delta$.
The red horizontal line shows the
gap $\Delta_{\rm TI}$,
induced in the topological insulator by the proximity effect. Note that the
state with
$\mu = 2$
has zero probability density at the center of the hole. Thus, it cannot be
observed in a tunneling spectrum if the probe is located at
$r=0$.
For
$R < R_{\rm cr} \approx 3$,
only the state with
$\mu=1$,
$n=0$
and the Majorana fermion contribute to the spectrum below the gap
$|\omega| < \Delta_{\rm TI}$.
}
\label{exited_states}
\end{figure}

\begin{figure}[t!]
\center
\includegraphics [width=8.5 cm, height=7 cm]{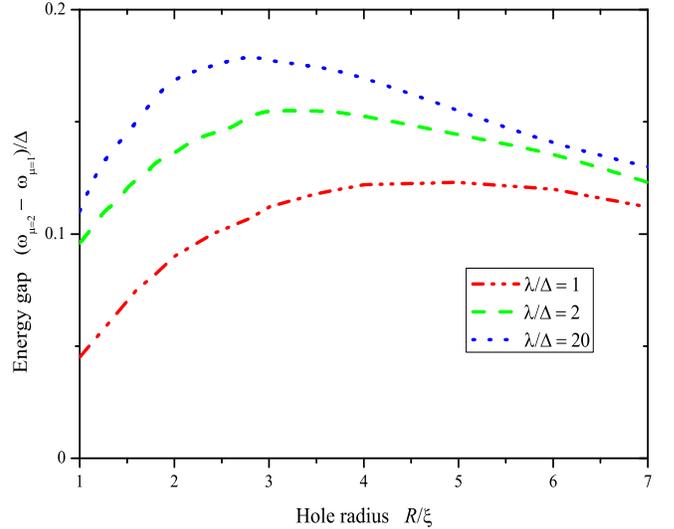}
\caption{(Color online) Energy gap between two low-lying excited states,
the first excited state
($\mu=1$, $n=0$)
and the state
$\mu=2$, $n=0$,
as a function of the normalized hole radius
$R/\xi$.}
\label{second_ex}
\end{figure}

For
$R/\xi \lesssim 4.5$
the second excited state has the quantum numbers
$\mu=2,\, n=0$
(see
Fig.~\ref{exited_states}).
With good accuracy, the energy difference between the first and second
excited states is
\begin{eqnarray}
\omega_{2}-\omega_{1} > 0.1 \Delta,
\end{eqnarray}
when
$R/\xi<4$, and $\lambda>2 \Delta$
(see
Fig.~\ref{second_ex}).
However, the probability density at the center of the hole,
$\rho_2 (0)$,
vanishes for this state. Thus, a tunneling experiment, in which the probe
is positioned near the center of the hole
($r=0$),
cannot detect this state, unless disorder is present.

The eigenenergies of the lowest-lying excited states are shown in
Fig.~\ref{exited_states}.
As one can see from this figure, when $R$ is smaller than some critical
value
$R_{\rm cr}$,
only two excited states remain. Of these two, only
$\mu=1$
state has finite probability density at
$r=0$.
For
$R>R_{\rm cr}$,
more states split off from the continuous spectrum and form bound states
inside the gap
$\Delta_{\rm TI}$.

\subsection{Back-of-the-envelope estimates}
\label{back-env_estimates}

The numerical results for the first excited state can be checked against
simple "back-of-the-envelope" calculations. A wave function of a subgap
state on the surface of the topological insulator is finite for
$r<R$,
but decays quickly for
$r-R > \xi$.
In other words, because of the superconducting gap, an electron with
energy
$|\omega_1| < \Delta$
is effectively confined to an area of radius
$R_{\rm conf} = R + \xi$.
Therefore,
$|\omega_1| \approx v |{\bf k}_1|$,
where the quantized momentum
$|{\bf k}_1| \approx \pi/(2R_{\rm conf})$.
This means that
\begin{eqnarray}
\label{box}
\frac{\omega_1}{\Delta} \, \approx \, \frac{\pi}{(2R/\xi) + 2}
\, \approx \,
\begin{cases}
0.8 &{\rm \ \ if\ } R/\xi=1.0, \\
0.2 &{\rm \ \ if\ } R/\xi=7.0.
\end{cases}
\end{eqnarray}
These numbers agree well with the numerical data for large barrier
transparencies (see
Fig.~\ref{first_ex}).
The quality of this estimate deteriorates for smaller $\lambda$, because in
this regime the induced gap decreases, and the confinement of the subgap
state becomes weaker. As a result, our simple estimate for
$R_{\rm conf}$
becomes inaccurate, at least for small $R$ (for larger $R$ the accuracy of
this estimate improves, since the hole radius becomes the dominant
contribution to
$R_{\rm conf}$).


\subsection{Resolving exited states}

In this section we will discuss the optimization of our system for the purpose
of resolving the excited states. We will assume that the STS tip is placed
above the center of the hole. In such a situation, only states with
$\rho(0)>0$
contribute to the tunnel current. As can be seen from
Eq.~\eqref{Center},
only states with
$\mu=0,\pm1$
have a non-zero probability density at the center of the hole. The
numerical analysis of
Eq.~\eqref{Energy_eq}
shows that, when
$|\mu| \leq 1$,
the lowest excited state corresponds to the quantum numbers
$n=0,\,\mu=1$,
next is the state
$n=1,\,\mu=0$,
and afterward
$n=2,\,\mu=0$.

When
$R < R_{\rm cr}$,
of these three states only the state with
$n=0$, $\mu = 1$
remains inside the gap. Two others are virtually merged with the continuum
spectrum above
$\Delta_{\rm TI}$.

A hole with a radius of
the order of
$R_{\rm cr}$
is optimal for the observation of the first excited state. Indeed, in this
regime only the first excited state contributes to the tunneling current at
the center of the hole. Furthermore, for a broad range of transparencies
$\lambda$, this state lies close to the middle of the gap
$\omega_1 \approx \Delta_{\rm TI}/2$,
being well-separated from both the Majorana state at
$\omega = 0$
and from the continuum at
$|\omega| = \Delta_{\rm TI}$.

\begin{figure}[t!]
\center
\includegraphics [width=9 cm, height=7 cm]{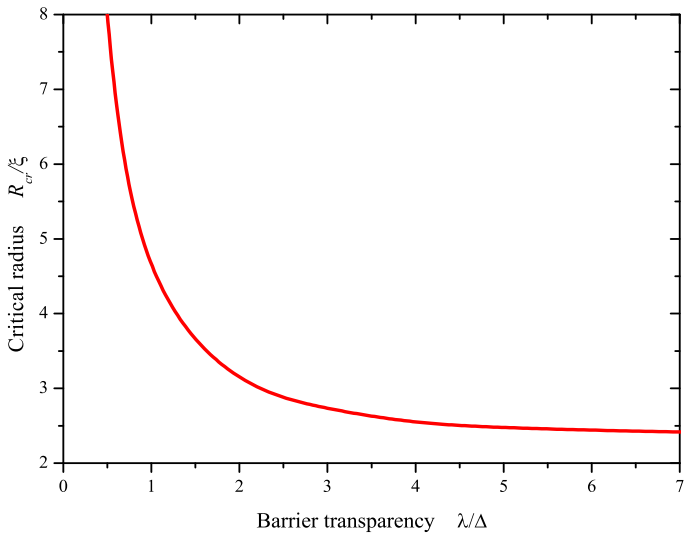}
\caption{Critical radius $R_{\rm cr}$ as a function of the normalized
transparency
$\lambda/\Delta$
of the barrier. The critical radius is defined by the requirement that for
$R<R_c$
only one subgap state has non-zero density of states at the center of
the hole.
}
\label{crit_R}
\end{figure}


The dependence of
$R_{\rm cr}$
on the barrier transparency $\lambda$ is shown in
Fig.~\ref{crit_R}.
The optimal transparency of the barrier
$\lambda \geq 2\Delta$,
thus, the optimal radius is
$2\xi<R<3\xi$.
For these parameters, the gap between the Majorana fermion and the first
excited state is about
$0.4\Delta$,
and between the first excited state and
the continuum is about $0.3\Delta$.

\section{Discussion}
\label{discussion}

Current interest in Majorana fermions is fueled, among other reasons, by
the possibility to devise a future topological quantum computer. To realize
this Majorana fermion-based computer, Majorana fermion localized states
must be created and moved in space in a controllable manner. At present,
this appears to be a very distant goal. The more modest objective of
creating an immobile Majorana fermion is being pursued now, and certain
initial steps are happening in this
direction.~\cite{frolov,Goldhaber}
However, no decisive proof of Majorana fermion states is available.
In this paper we study a simple
heterostructure~\cite{AL}
where an immobile Majorana fermion can be generated. Despite its relative
simplicity, the proposed system has several advantages, which can be useful
for the experimental detection of the Majorana state. In this section we
offer a nontechnical summary of the system's most important features.

\subsection{Large minigap}

One of the key characteristics of our system is the substantial energy gap
between the zero-energy Majorana fermion and the lowest excited state.
This is important since it alleviates requirements on the temperature and
the energy resolution of the experiment. By choosing the system parameters
adequately, the energy of the first excited state can be as large as
0.4$\Delta \sim 4$\,K.
This gap is much larger than the minigap for CdGM states
$\delta \varepsilon \sim \Delta^2/E_{\rm F} \sim 10^{-2}$\,K
(we assume that
$E_{\rm F} \sim 10^4$\,K
and
$\Delta \sim 10$\,K).

To understand the origin of such a large minigap in our system we can
resort to a simple "particle-in-a-box" estimate: A massless Dirac fermion
with energy
$\omega < \Delta_{\rm TI}$
is trapped inside a disk of radius
$R \sim \xi$
(the entrapment occurs because the particle energy is below the gap
$\Delta_{\rm TI}$,
thus, it cannot propagate in an environment with a gap, which exists for
$r>R$).
This simple estimate reproduces the numerical results quite accurately: see
Eq.~(\ref{box}).
Analyzing the derivation of
Eq.~(\ref{box}),
one concludes that the large minigap is a consequence of the linear
spectrum of the excitations on the surface of the topological insulator.

Finally, we would like to cite
Refs.~\onlinecite{weak_prox},
which studied similar heterostructures in the limit of the weak proximity
effect
$\lambda \ll \Delta$
(we did not study this regime, since it corresponds to a very low induced
gap:
$\Delta_{\rm TI} \ll \Delta$).
These papers established that the minigap is of the order of the
proximity-induced gap. How can these results be applied to our case where
$\lambda \sim \Delta$?
Note that for very weak $\lambda$ the minigap is an increasing function
of $\lambda$. When $\lambda$ becomes comparable to $\Delta$, the minigap
reaches some finite value
$\Delta^*$.
How does this value compares against $\Delta$? We notice that in the regime
$\lambda \sim \Delta_{\rm TI} \sim \Delta$
there is only one energy scale in our system, and we conclude that
$\Delta^* \sim \Delta$.

The system with such a large minigap deserves a detailed study. Here our
aim was twofold: to investigate how the local tunneling spectroscopy can be
used to prove the existence of the Majorana fermion in our setup
(Sec.~\ref{tunneling_spectr}),
and to optimize the parameters of the system for such an experiment
(Sec.~\ref{optimization}).

\subsection{Tunneling spectroscopy of the core}
\label{tunneling_spectr}

The Majorana fermion should manifest itself on a tunneling experiment as a
zero-bias peak. However, the zero-bias peak may be caused by other
mechanisms (see, for example, the analysis of
Ref.~\onlinecite{zbp}),
thus, additional verifications are necessary. In this paper we discussed
two types of further measurements. First, one can study the dependence of
the zero-bias anomaly on the magnetic field. As it follows from
Eq.~(\ref{condit}),
the number of zero modes is equal to the vorticity pinned by the hole. Thus,
when the field is increased, in the disorder-free system the strength of
the zero-bias anomaly should experience a stepwise increase each time an
extra flux quantum enters the pinning hole. If disorder is present, the
behavior of the zero-bias anomaly changes. The disorder potential lifts the
degeneracy of the zero-energy states (splitting of the zero-energy manifold
by a perturbation is studied in
Ref. 38).
However, due to symmetry
[see~Eq.~(\ref{charge_conj})],
the parity of the zero-mode number remains unchanged by the disorder.
Therefore, for even (odd) vorticity $l$ there is no (single) zero-energy
Majorana fermion state bound to the hole. This means that, if weak disorder
is present, the zero-bias anomaly demonstrates a non-monotonous dependence
on the magnetic field. The experimental verification of such a
non-monotonicity would be a strong argument in favor of Majorana fermion
states in our heterostructure. Of course, inducing multiply quantized
vortex in experiment is a complicated, but not insurmountable,
issue~\cite{roditchev}.

Magnetic field may also lift the degeneracy of the zero-energy states. We
already explained above that the magnetic field significantly affects a
particular state only when the flux through the area where this state is
localized is comparable with the flux quantum. In our situation, this
condition is not satisfied, and it is possible to apply perturbation theory
in orders of the vector-potential
${\bf A}$
to account for the magnetic field.
Equations~(\ref{r>R})
and
(\ref{r<R})
for the zero-energy wave functions valid for $U=0$ can be used to evaluate
the corresponding matrix elements. However, it is easy to check that these
matrix elements are identically zero. They may become finite only when
$U \ne 0$.
Consequently, in the limit
$|U| \ll \Delta$,
which is the most suitable for observation of the Majorana fermion, the
splitting due to the magnetic field is very weak, at least when the trapped
vorticity remains small. When the vorticity grows, a more advanced
treatment might be required. However, at large vorticity the detrimental
effects of magnetic field on the superconducting structure degrade the
performance of the system in a variety of ways. Thus, the limit of strong
magnetic field is outside the optimal regime, and we will not study it in
this paper.

The second type of measurements we discussed is the resolution of the
excited subgap states bound to the hole. Unlike the classical CdGM states,
which densely fill the core of a vortex, only a small number of subgap
excitations exists in our setup. Strictly speaking, a successful detection
of these excitations does not constitute a proof for Majorana fermion
existence. Yet, it would provide an additional check point validating the
theoretical description of the heterostructure.

\subsection{System parameter optimization}
\label{optimization}

\begin{figure}[t!]
\center
\includegraphics [width=8.5cm, height=6.5cm]{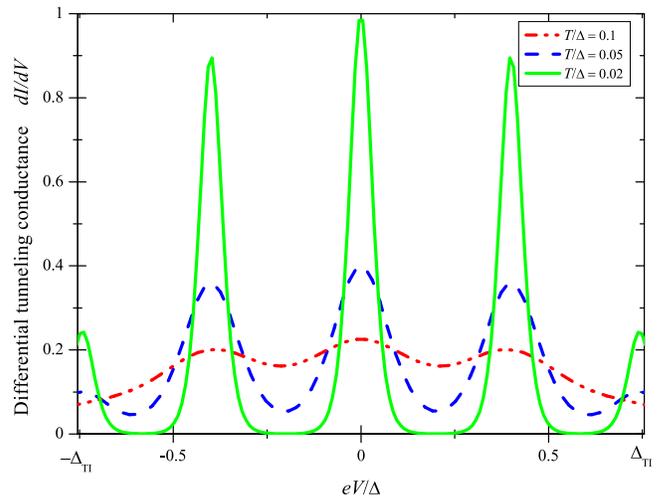}
\caption{
(Color online) Differential tunneling conductance for different
temperatures and optimal values of the parameters
$R/\xi=3$
and
$\lambda/\Delta=2$.
Majorana state is responsible for the zero-bias peak. When the STS tip is
placed above the center of the hole
($r=0$),
only one excited state
($\mu=\pm 1$, $n=0$)
contributes to the tunneling spectrum.
\label{tun_current}
}
\end{figure}

To facilitate experiments we investigated the possible optimization of the
system parameters. We found that if the tunneling probe is situated over
the center of the hole, only excitations with small $n$ and $\mu$
contribute to the tunneling conductance. Excited states with higher $n$ and
$\mu$ are localized closer to the hole periphery; thus, they do not affect
significantly such a tunneling spectrum. Consequently, if
\begin{equation}
\label{optimal}
\frac R{\xi}\  \simeq \  2\! -\! 3,\quad\quad
\frac {\lambda}{\Delta}\  \geq\  2,
\end{equation}
then only a single subgap state can be seen in the tunneling spectrum
measured at
$r=0$.
Under these conditions, the energy gap between the Majorana fermion and
this excited state is about
$0.4\Delta$,
and between the excited state and
the continuum above the superconducting gap is about
$0.3\Delta$.
The numerically calculated tunneling conductance for this situation is
presented in
Fig.~\ref{tun_current}.

To obtain some estimates, let us now take the characteristic values
$T_c=10$~K
and
$\Delta=1.76T_c \approx 17.6$~K
for a BCS-type superconductor. Then we obtain the optimal value
$\lambda\gtrsim 4$~meV.
From Fig.~\ref{tun_current}
we conclude that
$T\lesssim$~0.2--0.5~K is needed to resolve the Majorana fermion and the
excited states. To find the radius of the hole we use the formula
$\xi = v/\Delta \approx 200$~nm
(using the value
$v = 5.0 \times 10^7$~cm/sec
reported in
Ref.~\onlinecite{fermi_vel}
for Bi$_2$Se$_3$).
Thus,
$R \sim$~400--600~nm.

During our discussion we tacitly assumed that the coherence length in the
superconductor
$\xi_{\rm SC}$
is identical to
$\xi = v/\Delta$.
For BCS superconductor this implies that the Fermi velocity in the
superconductor is equal to $v$. Fortunately, such a restriction may be
replaced by a much weaker requirement:
$R > \xi_{\rm SC}$.
This guarantees that the vortex core contains no CdGM states, and our
derivation of the effective Hamiltonian is valid.

To conclude, we discuss the application of scanning tunneling spectroscopy
to investigate localized states in the topological insulator/superconductor
heterostructure presented in
Fig.~\ref{stm_exp}.
STS can be used to detect the oscillation of the zero-bias anomaly strength
when the magnetic field is varied, and to resolve subgap excited states.
The successful observation of both phenomena would provide strong evidence
in favor of the existence of a Majorana fermion state bound to the hole.

\section*{Acknowledgments}

This work was partly supported by the ARO,
RIKEN's iTHES program,
MURI Center for Dynamic Magneto-Optics,
Grant-in-Aid for Scientific Research (S),
MEXT Kakenhi on Quantum Cybernetics, the JSPS via its FIRST program,
the Russian Foundation for Basic Research (projects No. 11-02-00708,
No. 11-02-00741, No. 12-02-92100-JSPS, and
No. 12-02-00339).
The authors would like to thank S. V.~Zaitsev-Zotov for useful comments.

\end{document}